\documentclass[iop, apj]{emulateapj}

\usepackage{amsmath} 

\submitted{Accepted for publication in ApJ}

\shorttitle{Orbits in a tidally induced bar}
\shortauthors{Gajda et al.}

\begin{document}

\title{The orbital structure of a tidally induced bar}
\author{Grzegorz Gajda\altaffilmark{1,2}, Ewa L. {\L}okas\altaffilmark{1} and E. Athanassoula\altaffilmark{2}}
\affil{
 \altaffilmark{1}Nicolaus Copernicus Astronomical Center, Polish Academy of Sciences, Bartycka 18, 00-716 Warsaw, Poland \\
 \altaffilmark{2}Aix-Marseille Universit\'{e}, CNRS, LAM (Laboratoire d'Astrophysique de Marseille) UMR 7326, 13388 Marseille, France
}

\begin{abstract}
Orbits are the key building blocks of any density distribution and their study helps us understand the kinematical structure and the evolution of galaxies. Here we investigate orbits in a tidally induced bar of a dwarf galaxy, using an $N$-body simulation of an initially disky dwarf galaxy orbiting a Milky Way-like host. After the first pericenter passage, a tidally induced bar forms in the stellar component of the dwarf. The bar evolution is different than in isolated galaxies and our analysis focuses on the period before it buckles. We study the orbits in terms of their dominant frequencies, which we calculate in a Cartesian coordinate frame rotating with the bar. Apart from the well-known x$_1$ orbits we find many other types, mostly with boxy shapes of various degree of elongation. Some of them are also near-periodic, admitting frequency ratios of 4/3, 3/2 and 5/3. The box orbits have various degrees of vertical thickness but only a relatively small fraction of those have banana (i.e. smile/frown) or infinity-symbol shapes in the edge-on view. In the very center we also find orbits known from the potential of triaxial ellipsoids. The elongation of the orbits grows with distance from the center of the bar in agreement with the variation of the shape of the density distribution. Our classification of orbits leads to the conclusion that more than $80 \%$ of them have boxy shapes, while only $8 \%$ have shapes of classical x$_1$ orbits.
\end{abstract}


\section{Introduction}
\subsection{Bars in the Universe}

Bars are characteristic components of many disk galaxies.
Their abundance estimates vary, depending on the criteria and method employed,
the fraction of barred galaxies ranging from $25$ to $60 \%$ of disk galaxies \citep[and references therein]{masters11}.
For bright dwarf galaxies in the Virgo cluster, \citet{janz12} found the bar fraction of $18\%$.

Bars can be formed via an instability occurring in a stellar disk.
In the beginning, the bar grows very fast, but soon after it buckles out of the disk plane \citep{combes90, raha91}.
Later on, it experiences a secular evolution phase, lengthening gradually and slowing down, i.e. decreasing its pattern
speed $\Omega_\mathrm{p}$. For reviews on the subject of bars, see \citet{sellwood_wilkinson93} and
\citet{athanassoula13}.

Bars can also be formed through tidal interactions of large disk galaxies \citep{noguchi87,
gerin90}, either as a result of an encounter with a smaller perturbing galaxy
\citep[e.g.][]{lang14} or perturbations from a galaxy cluster \citep{lokas16}. However, it is not
entirely clear whether the bars formed in this way have the same dynamical properties as bars
formed in isolation. For example, \citet{noguchi96} argued that they have different shapes
of the density profiles. On the other hand, \citet{berentzen04} claimed that bars formed in isolation and in
interactions have the same or similar dynamical properties and that this made it very difficult (if at all possible) to
distinguish between the two formation channels. Most of the studies agree that the bars formed tidally have smaller
pattern speeds \citep{miwa_noguchi98, berentzen04, lokas16}. In addition, \citet{miwa_noguchi98} claimed that the
tidally induced bars end at the Inner Lindblad Resonance, whereas the bars formed in isolation extend up to roughly
$80\%$ to $100\%$ of the Corotation Resonance \citep{athanassoula92a, athanassoula92b}. Moreover, the properties of the
bars formed via interactions seem to depend on the tidal force strength \citep{miwa_noguchi98,lokas16}. If the
interaction is weak, then the tidal force merely induces the instability and the bar parameters are set by the
properties of the disk, whereas for the stronger interaction, its properties are determined by the encounter.

In the tidal
stirring scenario for the formation of dwarf spheroidal galaxies, initially disky dwarfs are captured by larger
hosts (like the Milky Way) and transformed into spheroids by the tidal forces \citep{mayer01,
klimentowski09, kazantzidis11, lokas11}. An intermediate stage of this process is the formation of a bar in the dwarf
galaxy, which should still exist and be observable in some dwarf spheroidals of the Local Group. Indeed, good
candidates for such bar-like systems exist, including the Sagittarius, Ursa Minor and Carina dwarfs \citep{lokas10,
lokas12}.
The elongated shapes of some recently discovered ultra-faint dwarf galaxies, like Hercules
\citep{coleman07} and Ursa Major II \citep{munoz10}, also indicate possible presence of bars.

Recently, \citet{lokas14} analyzed an $N$-body simulation of a dwarf galaxy on an elongated, prograde orbit
around a Milky Way-like host. In this simulation, the bar formed after the first pericenter passage
and during subsequent encounters with the host it was shortened and weakened. It also experienced a buckling
episode after the second pericenter. Its properties changed abruptly at pericenter passages but were relatively stable
between them. In a follow-up study, \citet{lokas15} demonstrated that the efficiency of bar formation
depends very strongly on the initial inclination of the dwarf galaxy disk and the bar does not form in exactly
retrograde configurations.

\subsection{Orbits in bars}

Early studies of the orbital structure of bars concentrated on periodic orbits in potentials resembling observed
galaxies. \citet{contopoulos_papayannopoulos80} described four main orbital families in 2D bars. The x$_1$ orbits
are elongated in the same direction as the bar. At their ends they may have cusps or even small loops. Two other families
are perpendicular to the bar: x$_2$ orbits are stable, while x$_3$ are unstable. The last family identified included
orbits named x$_4$, which are not far from circular and, importantly, retrograde with respect to the bar rotation. Stars in a given
potential should mainly follow orbits with shapes contributing to the density distribution generating the
potential. Thus, it was argued that bars should be mostly made of x$_1$-related orbits
\citep{athanassoula83}. This finding was afterwards confirmed by a 2D $N$-body simulation performed by
\citet{sparke_sellwood87}.

Later analyses studied the impact of the third (vertical) dimension on the orbital structure.
\citet{pfenniger_friedli91} noted that the main 3D orbits contributing to their $N$-body bar, when viewed edge-on, were banana-shaped or had a shape of the infinity symbol. They were named {\sc ban} and {\sc aban} orbits, respectively, as an extension of the \citet{athanassoula83} notation, where the x$_1$ orbits were named as B.
An exhaustive classification of non-planar orbits was performed by \citet{skokos02}, who extended the notation of \citet{contopoulos_papayannopoulos80} to three dimensions and thus introduced the x$_1$v$_1$, x$_1$v$_2$ etc. families.

One of the methods to study particle orbits relies on the analysis of dominant frequencies \citep{binney_spergel82,
laskar90, carpintero_aguilar98}. It is based on finding peaks in the Fourier spectrum of a given coordinate time series. If two frequencies
are commensurate, then in a given reference frame the orbit is periodic, i.e. it will close after some time and
repeat itself.

In the case of bars, following \citet{athanassoula02}, the orbits are usually described in terms of the epicycle
frequency $\kappa$ and the angular frequency $\Omega$. When the vertical structure is also studied, the vertical
frequency is added. As was found by many studies \citep[e.g.][]{athanassoula02, athanassoula03,
martinez_valpuesta06, ceverino_klypin07}, the highest peak in the distribution of frequency ratios occurs at $(\Omega-\Omega_\mathrm{p})/\kappa=1/2$.
The proposed interpretation of this fact was that those orbits are x$_1$-related, i.e. during one orbital period they experience two radial oscillations.

In most of the studies of orbits performed to date the orbits were analyzed in a frozen potential. A common approach is
to extract a snapshot from an $N$-body simulation, calculate the (often symmetrized) potential and let it rotate with
the pattern speed of the bar. Then the trajectories of selected particles are followed in such a non-evolving setting.
One of the few studies adopting a different approach was performed by \citet{ceverino_klypin07} who studied actual
particle orbits in an ongoing $N$-body simulation.
This leads to a considerably worse resolution in comparison with using a frozen potential because of the
ubiquitous secular evolution, but this resolution is sufficient for our purposes. However, their results
were in broad agreement with previous works.

Knowledge of the orbital structure of a galaxy can be useful in many ways. Orbits are building blocks of the density
distribution and are used in the \citet{schwartzschild79} modelling of barred galaxies \citep{zhao96,
vasiliev_athanassoula15}. The high velocity feature in the Milky Way bar \citep{nidever12} is claimed to be explained
by the contribution from some particular orbit types \citep{aumer_schonrich15, molloy15}. The gas flow in barred
galaxies is also directly related to the periodic orbits \citep{binney91, athanassoula92b, sormani15}.

All the above works, however, pertain to bars in galaxies of Milky Way type, while no study so far addressed the
orbital structure in barred dwarf galaxies. In this paper we extend the work of \citet{lokas14} by an in-depth study
of stellar orbits in the tidally induced bar using the dominant frequency analysis. Our purpose is to establish
whether the orbital structure of such a bar is similar to the one of previously studied bars formed in isolation.
In section \ref{sec_methods} we characterize the simulation used and describe the details of our frequency analysis
method. In section \ref{sec_results} we present the results, starting with an overview of shapes of the individual
stellar orbits. Then, we analyze a large number of orbits and study their properties. Next, we classify the orbits in
terms of their shapes and frequency ratios and quantify their contribution depending on the distance from the center of
the dwarf. We finish this section with the description of orbits which buckle out of the disk plane. Finally, in
section \ref{sec_discussion} we discuss our results and compare them to previous works. The paper is concluded with a
summary in section \ref{sec_summary}.

\section{Methods}
\label{sec_methods}
\subsection{The simulation}
We reanalyzed the $N$-body simulation of an interaction between a dwarf galaxy and its host, previously described in
\citet{lokas14}. It employed a Milky Way-like galaxy model, made of an NFW \citep{nfw95} dark matter halo and an
exponential stellar disk. The halo had mass $7.7 \times 10^{11}$ M$_\sun$ and concentration $27$. The disk had mass
$3.4 \times 10^{10}$ M$_\sun$, the radial scale-length of $2.82$ kpc and thickness of $0.44$ kpc. The model of the dwarf
galaxy was also two-component. Its NFW halo had mass $10^9$ M$_\sun$ and concentration $20$. The dwarf's disk had mass
of $2 \times 10^7$ M$_\sun$, the radial scale-length of $0.44$ kpc and thickness of $0.088$ kpc. Each galaxy was
built as an equilibrium $N$-body realization of $10^6$ particles per component ($4 \times 10^6$ particles in total).
The dwarf was initially placed at the apocenter of an elongated, prograde orbit, with the apocenter distance of
$120$~kpc and pericenter of $24$ kpc.

The simulation, followed for $10$ Gyr, was performed using publicly available code \textsc{gadget2} \citep{springel05}.
The adopted softening lengths were, respectively, $2$ kpc and $0.05$~kpc for the halo and the disk of the host galaxy,
and $0.06$ kpc and $0.02$ kpc for the halo and the disk of the dwarf. We rerun the simulation and saved outputs every $0.005$ Gyr,
i.e. ten times more frequently than in \citet{lokas14}.

As already mentioned, in this simulation the initially axisymmetric disk of the dwarf galaxy forms a bar after the
first pericenter passage. During subsequent encounters with the host, the bar gets shorter and thicker. For the purpose
of the present study, it seems to be most appropriate to consider the bar when it is the strongest and for a period
of time when it remains approximately unaltered. Thus, we chose the period between the first ($1.2$ Gyr) and the second
($3.3$ Gyr) pericenter passage.

\begin{figure}
\centering
\includegraphics[width=\columnwidth]{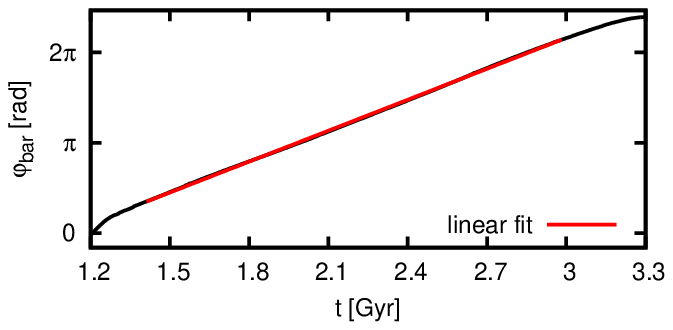} \\
\includegraphics[width=\columnwidth]{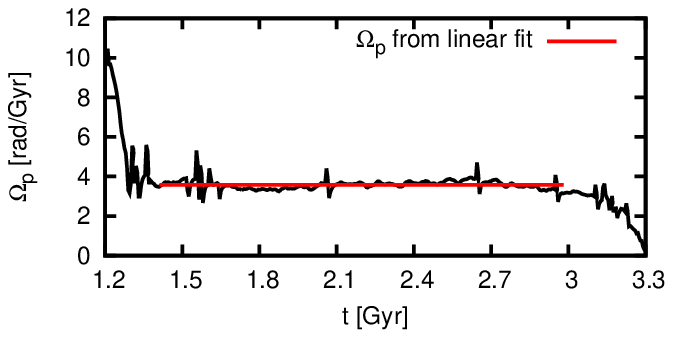}
\caption{Top: position angle of the major axis of the bar (black) with the linear fit in the time interval 1.4-3.0
Gyr (red). Bottom: the pattern speed of the bar measured from the simulations (black) and the best-fitting value
inferred from the slope of the linear fit to the position angle (red).}
\label{fig_major_axis}
\end{figure}

\begin{figure*}
\centering
\includegraphics[width=\textwidth]{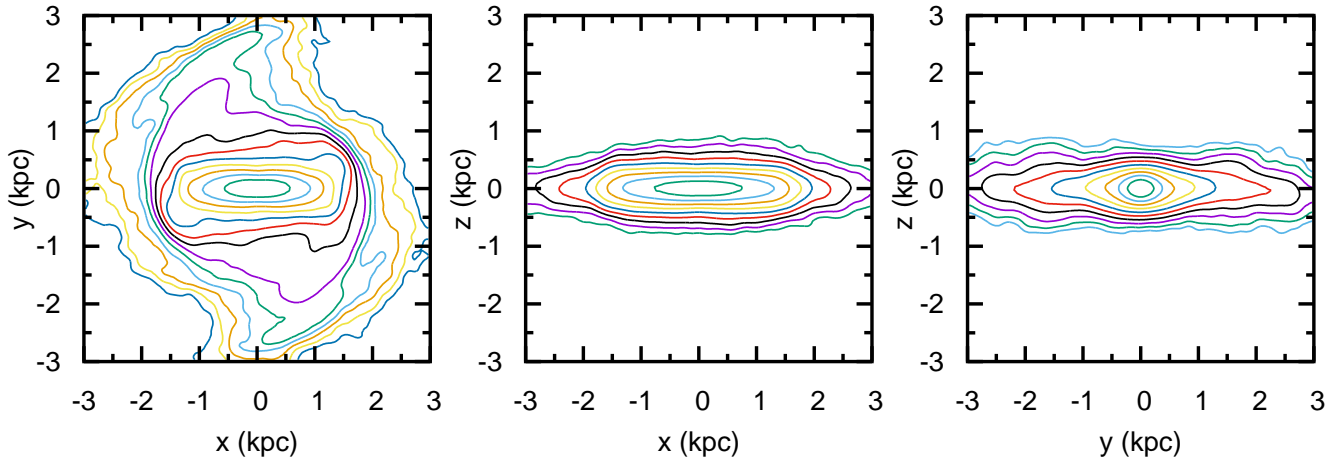}
\caption{The surface density contours of the stellar component of the dwarf galaxy at $2.2$ Gyr, when it reached
the second apocenter of its orbit. From the left to the right: the face-on, edge-on and end-on views.
The contours are spaced logarithmically: the outermost one corresponds in all panels to the surface density of $5.6\times10^3$ M$_\sun/\mathrm{kpc}^2$. In the left panel, each subsequent contour corresponds to a density $1.88$ times larger, whereas in the middle and right panels the contour spacing has a multiplier of $2.67$.}
\label{fig_density_maps}
\end{figure*}

In order to study orbits in an environment that changes as little as possible with time, we determined the
period when the bar rotates with an approximately constant pattern speed. For this purpose, for each simulation output
we calculated the principal axes and axis ratios from the shape tensor, using the method of \citet{zemp11}
as implemented by \citet{gajda15} and taking into account the stellar particles in an ellipsoid of mean radius equal to
$0.5$ kpc. During the studied period, the disk is tilted by a few degrees with respect to the initial plane and
precesses. In order to correct for this, we had to find a time-averaged disk plane.

In the top panel of Figure \ref{fig_major_axis} we plot the position angle $\varphi_\mathrm{bar}$ of the bar major
axis. In the time interval of 1.4-3.0 Gyr it is well approximated by a linear function
$\varphi_\mathrm{bar}(t)=\Omega_\mathrm{p} \, t + \varphi_\mathrm{bar} (t=0)$. In the bottom panel of Figure
\ref{fig_major_axis} we show the pattern speed $\Omega_\mathrm{p}$ calculated from the simulation, along with the
one obtained from the fit. In the considered period of time the pattern speed fluctuates a little, but is fairly
constant, therefore we adopt the fitted value as the pattern speed of the bar.

We note that a constant pattern speed is not a common feature of bars formed in isolation. Usually,
the bar slows down during its evolution; initially it decelerates strongly and later on, in the secular evolution phase, the slowdown is more gradual \citep[for a review]{athanassoula13}.
The lack of detectable bar slowdown could well argue for a different angular momentum redistribution than in isolated galaxies, which may involve an angular momentum exchange with the host galaxy.
For a bar formed through interaction \citet{miwa_noguchi98} also found that the pattern speed seems to be rather constant in time.

Figure \ref{fig_density_maps} shows surface density maps of the stellar component of the dwarf galaxy after 2.2 Gyr of
evolution, when the dwarf is at the second apocenter of the orbit. The coordinate system used in this Figure will be
used throughout the paper: the $x$ axis is aligned with the bar major axis, the $y$ axis with the intermediate and $z$
with the minor one, perpendicular to the disk. In the face-on view the bar is clearly visible in the center. At some
distance the contours begin to twist and the disk starts to dominate while at the outskirts the tidal tails are
discernible. The bar has not
yet undergone the buckling instability, hence it does not have a boxy/peanut shape in the edge-on view.

\begin{figure}
\centering
\includegraphics[width=\columnwidth]{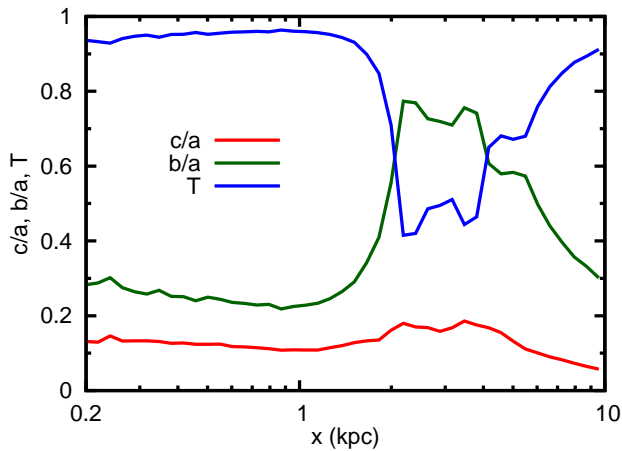}
\caption{The axis ratios of the stellar component of the dwarf galaxy: shortest to longest ($c/a$, red) and
intermediate to longest ($b/a$, green), supplemented with the triaxiality parameter ($T$, blue) for the bar
at the apocenter of the dwarf's orbit (at $2.2$ Gyr). All quantities are shown as a function of distance from the
dwarf center measured along its major axis.}
\label{fig_shape_x}
\end{figure}

In the following, we are going to study the orbits as a function of their distance from the center of the dwarf. Thus,
it would be useful to know if there are any changes of the bar shape with radius and how long the bar is.
Using the method mentioned before, we calculated the axis ratios as a function of distance measured along the bar major axis. In
Figure \ref{fig_shape_x} we plot the ratios of the minor to major axis ($c/a$), intermediate to major axis ($b/a$) and
the triaxiality parameter $T=[1-(b/a)^2]/[1-(c/a)^2]$. If $T<1/3$, the object can be considered oblate, if
$1/3<T<2/3$ it is triaxial and finally if $T>2/3$ the stellar distribution is prolate, i.e. bar-like.

The $c/a$ ratio, i.e.\ the thickness of the galaxy, is approximately constant over the whole dwarf. On the other hand,
the $b/a$ ratio varies significantly. It decreases slowly in the inner parts but then increases rapidly, which
signifies the transition to the disk. Further away it drops again as we move from the main body of the galaxy to the
tidal tails. These changes of $b/a$ are reflected in the variation of the $T$ parameter. In the inner part, dominated
by the bar, $T$ is almost constant (or even grows slightly) and is of the order of $0.95$. Further away it falls below
$2/3$, which means that the dwarf's shape becomes triaxial at this radius. At large distance, $T$ grows again above
$2/3$, indicating the transition to the tidal tails.

Unfortunately, there is no unique method to measure the bar length and the existing ones give
various estimates \citep{athanassoula_misiriotis02}. We base our estimate of the bar length on the
behavior of $T$ (for a similar approach based on ellipticity and other methods see e.g.
\citealt{athanassoula_misiriotis02}). When $T$ decreases, it crosses the value of $0.9$ at
$r\approx 1.6$ kpc, and the shape stops being triaxial ($T<2/3$) at $2$ kpc. We also checked how
the orientation of the major axis of the stellar density distribution changes with radius. Its
direction turns out to be very stable up to $1.6$ kpc, and only then starts to change slowly so
that at $2$ kpc it is significantly different than in the center. Given that both methods are
arbitrary in choosing the thresholds, we decided to adopt as the length of the bar semi-major axis
the mean of the low and high estimates, that is $1.8$ kpc. We verified that the bar length
estimated in this way does not change significantly over the period (between the first and second
pericenter) we have selected for our study of orbits. We also note that the bar length is always
shorter than the tidal radius at the corresponding time as calculated in \citet{gajda_lokas16}.

\subsection{Dominant frequencies}

We use an algorithm similar to the numerical analysis of fundamental frequencies (NAFF), developed by
\citet{laskar90} and discussed extensively by e.g. \citet{valluri_merritt98}. In the computations
we use $320$ simulation outputs from the period when the bar rotates steadily and proceed as
follows.

To study the orbits, we have to find the peaks of the Fourier transform (FT) of a time series of a given coordinate,
e.g.\ y(t). The FT $P_y(\omega)$ has the following form:
\begin{equation}
P_y(\omega)=\left| \frac{1}{t_2-t_1}\int\limits_{t_1}^{t_2} y(t)\exp[-i\omega (t-t_1)] \mathrm{d}t\right|,
\end{equation}
where $\omega$ is a frequency, and $t_1$ and $t_2$ are the beginning and the end of the period of interest. Because
$y(t)$ originates from a simulation, we have it in a sampled form as $y_n=y(t_1 + n\delta_t)$, where $n=0, 1, \ldots,
N-1$ and $\delta_t=(t_2-t_1)/N$. In our case $N=320$ and $\delta_t=0.005$ Gyr. Hence, we have to approximate the
integral as the sum:
\begin{equation}
P_y(\omega)=\left|\frac{1}{N} \sum\limits_{n=0}^{N-1} y_n \exp(-i\omega n \delta_t ) \right|.
\end{equation}
Let us note that we do not use any window function (as e.g. \citealt{laskar90}), because it widens the peaks. They are
already quite wide due to the small value of $N$ and broadening them further would result in overlapping of the
peaks in some spectra. $P_y(\omega)$ is a function of a continuous variable $\omega$ and initially we have no knowledge
of its shape. Hence, we calculate its values for a discrete set of frequencies $\{\omega_k\}$, which are equidistant
with $\delta_\omega=1$ Gyr$^{-1}$. We note that this spacing is a few times denser than in the case of the Fast Fourier
Transform (FFT) algorithm. We use a denser sampling since the one in the FFT is of the order of the width of an
individual peak and we found it insufficient. 

To find the maximum, we search for a maximal value of $P_y(\omega_k)$ and we denote its location by $\omega_K$. The
frequency $\omega_K$ is only an approximate location of the peak. Hence, following the Brent algorithm \citep{brent73,
numericalrecipies}, we search for the true maximum of the $P_y(\omega)$ function in an interval $(\omega_{K-p},\
\omega_{K+p})$, where $p$ is a small integer. We denote the obtained maximum of $P_y$ by $\omega_\mathrm{max}$.

However, the initial search for the maximum using $P_y(\omega_k)$ poses a risk. Let us imagine that the spectrum was
composed of two peaks: one exactly at $\omega_K$ and the other, slightly higher, at $\omega_L+\delta_\omega/2$
($\omega_K, \omega_L \in \{\omega_k\}$). When calculating the sampled spectrum, $P_y(\omega_K)$ of the first peak would
be picked exactly, but the second one would be probed only as $P_y(\omega_L)$ and $P_y(\omega_{L+1})$, both smaller
than $P_y(\omega_L+\delta_\omega/2)$. Hence, its height would be underestimated and the first peak, smaller in reality,
would be used for the refined search with the Brent algorithm.

\begin{figure}
\centering
\includegraphics[width=\columnwidth]{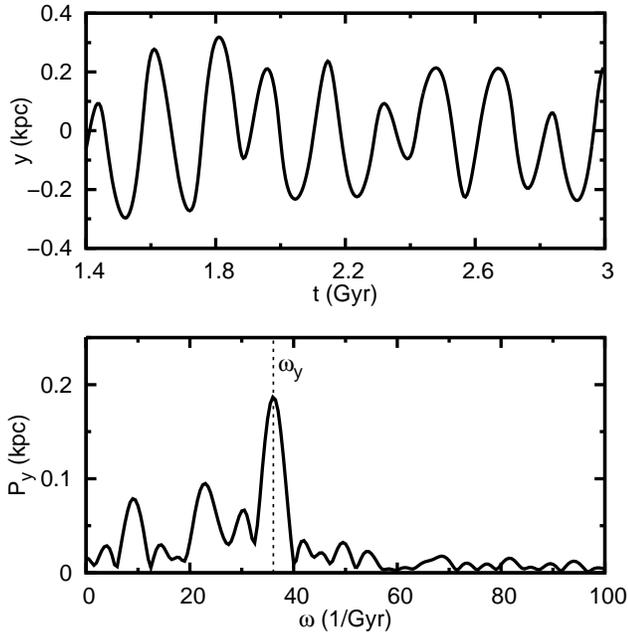}
\caption{Top: the variability of one of the coordinates in time, $y(t)$, for one of the stellar particles.
Bottom: the corresponding Fourier spectrum of $y(t)$. The frequency $\omega_y$ is indicated with a dashed line. The second highest
peak corresponds to $\omega_x$.}
\label{fig_freq_anal}
\end{figure}

To overcome this problem, we subtract a sine wave corresponding to the previously
found $\omega_\mathrm{max}$ from the initial time series $y(t)$, obtaining a transformed time series $y'(t)$. Then, we
search for the maximum of FT of $y'(t)$, using the same method as described above, and denote the position of the
maximum as $\omega'_\mathrm{max}$. Next, we search for the peak around $\omega'_\mathrm{max}$ in the FT of the
\emph{initial time series $y(t)$}. We compare the amplitudes (calculated from FT of $y(t)$) of the two peaks at
$\omega_\mathrm{max}$ and $\omega'_\mathrm{max}$ and we denote the frequency of the higher one as $\omega_y$. An
example of the time series $y (t)$ and its Fourier spectrum is shown in Figure \ref{fig_freq_anal}. In addition to the
highest peak, other significant lines are present in the spectrum. In particular, in the case of the $y$ spectrum, the
second highest peak usually has the same frequency as $\omega_x$.

The smaller peaks in the spectrum are also important and we retrieve the three most pronounced lines from each spectrum.
We subtract the mode with the estimated frequency $\omega_y$ from the initial time series and repeat the procedure
looking for next frequencies. Unfortunately, this procedure is still not unambiguous. Due to the subtraction of
the first peak, the second one may grow and in some rare cases become larger than the supposed highest one. Thus, in
the end we sort the lines. These issues point out to a major problem: if there are many lines in the spectrum, deciding
which one is dominant may be very difficult in some cases.

The width of a peak in the spectrum can be approximated as $\Delta\omega=2\pi/ T$, where $T$ is the total time over
which the Fourier transform is calculated. In our case $\Delta\omega\approx 3.93$ Gyr$^{-1}$.
First, this means that higher frequencies are determined relatively more accurately than lower ones. Second,
if $\omega_x\approx 12$ Gyr$^{-1}$ and $\omega_y\approx 16$ Gyr$^{-1}$, these two frequencies are barely resolved
in the $y$ spectrum. Therefore, for very low $\omega_y$ its determination becomes impossible. Moreover, at such low
frequencies the particles complete too few orbits, which influences the relative height of the peaks.

Usually, the $(x, y$) projection of orbits in bars is studied in terms of a ratio $(\Omega -
\Omega_\mathrm{p})/\kappa$, where $\Omega$ is the angular frequency, $\kappa$ is the epicyclic
frequency and $\Omega_\mathrm{p}$ is the pattern speed. These frequencies are calculated in a
cylindrical reference frame, centered on the galaxy center, hence we will refer to them as cylindrical
frequencies. Here, however, we chose to work in terms of the frequencies of the oscillations along
the $x$ and $y$ axes of a coordinate system rotating with the bar, which we will refer to as
Cartesian frequencies. There are several reasons for our choice. Given the shape of the bar, this
choice may seem more natural. We also wish to identify the various frequencies and families in the
same way as in triaxial ellipsoids, i.e. distinguish boxes from tubes (long-, or short-axis) and
not specifically in terms of the periodic orbit approach typically used for disk galaxies
\citep{athanassoula83, contopoulos_grosbol89}. We also do not wish to make comparisons with
specific theoretical work (e.g. to calculate angular momentum redistribution within the galaxy),
nor are we interested in the corotation resonance (i.e. the resonance in which
$\Omega=\Omega_\mathrm{p}$), which in fact is well outside the region we can study here. Last but
not least, the Cartesian frequencies are more straightforward to calculate than the cylindrical
ones. More specifically, it is the angular frequency $\Omega$ that is much more complex to
calculate and in many cases a straightforward application of a Fourier transform may lead to
erroneous results and should be supplemented with further techniques (\citealt{athanassoula02} and
in prep.; \citealt{ceverino_klypin07}). Discussing this is beyond the limits of this study and will
be the subject of a separate paper.

\begin{figure}
\centering
\includegraphics[width=\columnwidth]{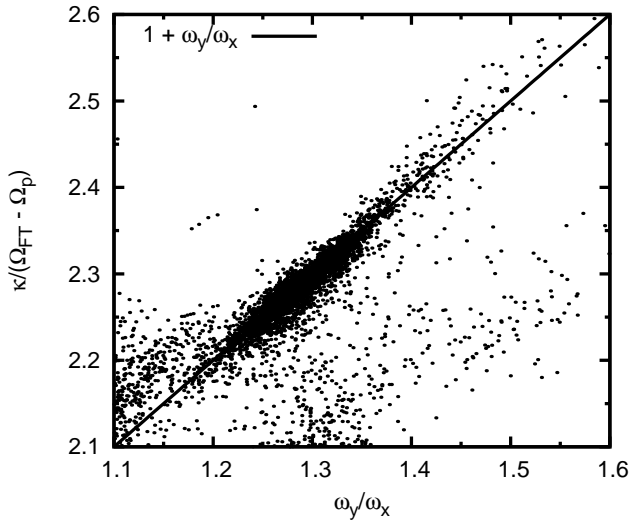}
\caption{A subsample of particle orbits admitting $(\Omega -\Omega_\mathrm{p})/\kappa<0.475$, i.e. those not
supporting the bar. The majority of the particles exhibit a correlation between
$(\Omega -\Omega_\mathrm{p})/\kappa$ and $\omega_y/\omega_x$.}
\label{fig_group_AB}
\end{figure}

Notwithstanding these issues, we made some experiments with the cylindrical frequencies in order to be able to compare
to previous results. For $\Omega$ calculated in a simplified way, namely from the Fourier transform of the particle
position angle $\varphi(t)$, we obtained that about $85\%$ particles analyzed had $(\Omega -
\Omega_\mathrm{p})/\kappa=1/2$. These orbits have $\kappa=2\omega_x$ and $\Omega - \Omega_\mathrm{p}=\omega_x$, which
is easy to understand and gives the $1/2$ ratio. Moreover, we found a considerable number of particles having $(\Omega
- \Omega_\mathrm{p})/\kappa \approx 0.44$ \citep[for a possibly similar group, see e.g.][]{harsoula09}. An inspection
of their shapes revealed that they are usually not quite elongated with the bar. They have more square shapes or are
even perpendicular to the bar. It turns out that the highest peak in their Fourier  spectrum of $r(t)$ no longer has
$\kappa = 2\omega_x$, but rather $\kappa = \omega_x + \omega_y$. Therefore, one can expect $\kappa/(\Omega -
\Omega_\mathrm{p}) = (\omega_x + \omega_y)/\omega_x = 1+\omega_y/\omega_x$. To illustrate this, we chose a subset of
particles having $(\Omega - \Omega_\mathrm{p})/\kappa < 0.475$ and plotted their frequency ratios in Figure
\ref{fig_group_AB}. It turns out that most of the particles follow this relation.

We thus perform our frequency analysis using $\omega_x$, $\omega_y$ and $\omega_z$, as well as $\kappa$.
We also retrieve the amplitudes of the Fourier peaks and denote them by $A_x$, $A_y$ etc. Some caution concerning the accuracy of our frequency determination is in order. Usually, tens of
orbital periods are considered to be required to reliably determine the frequencies. Here, we trace the orbits for
less than a dozen periods, so we performed two tests of accuracy for the particles having a mean distance from the potential center of $0.5$ kpc. First, we generated mock time series of $x$ and
$y$ coordinates. The time series were sums of three sine waves of amplitudes obtained for real particles, but their
phases were randomized. Then we used our method to recover the frequency ratios. For the x$_1$ orbits the dispersion
was of the order of $0.5 \%$. For the box orbits, it was larger and up to $1 \%$. It turned out that the
main contribution to the error comes from the presence of more than one peak in the Fourier spectrum. Thus this test may be considered simplified since we did not take into account e.g. harmonics. Second, as
x$_1$ orbits are expected to have the frequency ratio of $\omega_y/\omega_x=1$, we checked how accurately we recover
this value for the orbits from the simulation. We found the dispersion to be around $0.8\%$, slightly larger
than in the experiments with the mock data. Hence, we estimate that the overall accuracy of our frequency ratios is of
the order of $1.5\%$.
Unfortunately, in some cases an orbit has two peaks of similar amplitude in the spectrum.
Additionally, if such an orbit is integrated for too short a time, then the relative height of the
peaks will be changed. As a result, a misidentification of the highest frequency is possible. To
estimate how often this occurs, we would have to study subdominant frequencies and numerical noise, which is beyond the scope of this paper.

\section{Results}
\label{sec_results}
\subsection{Orbit shapes}

In this section we discuss the most common shapes of particle orbits and their characteristic frequency
ratios. First we consider face-on views of the orbits, then we proceed to {\sc (a)ban} shapes in the edge-on view.
Finally, we show the orbits found in the very center of the bar.

\begin{figure}
\centering
\includegraphics[width=\columnwidth]{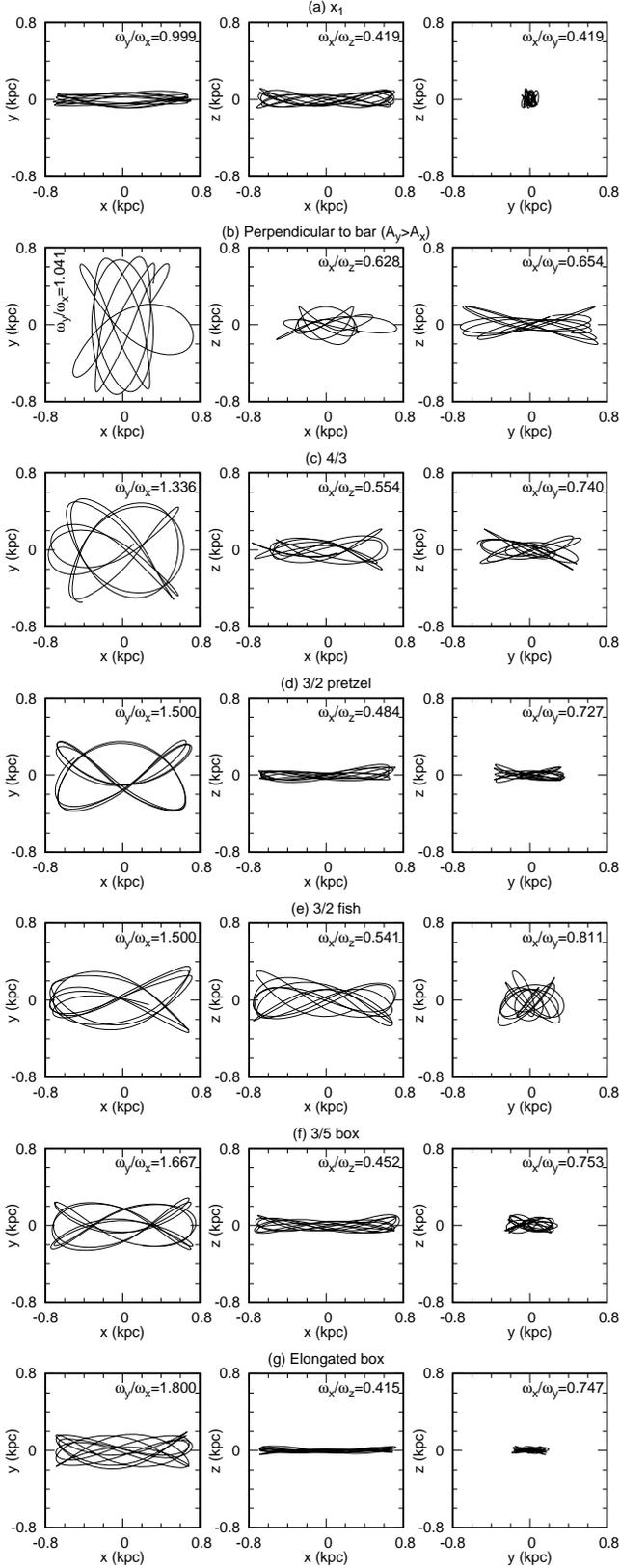}
\caption{Face-on, edge-on and end-on views of orbits, chosen by their frequency ratio and their shape in the face-on
view. The corresponding frequency ratios are indicated in the panels. All these orbits have the average size of
$\bar{r}=0.5$ kpc.}
\label{fig_orbits_flat} \end{figure}

In Figure \ref{fig_orbits_flat} we present various shapes of orbits in the plane of the disk, arranged according to the
growing $\omega_y/\omega_x$ ratio. In each row there are three projections of the orbits: face-on, edge-on and end-on.
The corresponding frequency ratios are indicated in the panels.

Row (a) of the Figure shows an x$_1$ orbit, which has $\omega_y/\omega_x\approx 1$.
We also found versions of this orbit with two loops at both ends, but then such orbits may have $\omega_y/\omega_x\approx
3$, depending on the size of the loops. Some have only one well developed loop and in those cases
$\omega_y/\omega_x\approx 2$. When viewed edge-on, they are usually flat.

In row (b) we depicted an orbit perpendicular to the bar for most of the time, i.e. $A_y>A_x$. Such orbits often have a
loop along the $x$-axis and are vertically extended, but not necessarily. They can be found approximately in a range
$1.05 < \omega_y/\omega_x < 1.15 $ and may be related to the classical families x$_2$ or x$_3$.

In row (c) we present an almost periodic orbit of $\omega_y/\omega_x\approx 4/3$. Judging by the face-on
plot such orbits do not appear elongated with the bar. Consequently, their strongest line in the $r(t)$ Fourier
spectrum is $\kappa=\omega_x+\omega_y$ and $(\Omega-\Omega_\mathrm{p})/\kappa\approx 0.44$ (see
Figure~\ref{fig_group_AB}). The vast majority of them have $\omega_x/\omega_z > 0.5$, thus they are vertically extended.
Taking everything into account, this type of orbits does not seem to support the bar.

In rows (d) and (e) we present two types of periodic orbits having $\omega_y/\omega_x\approx 3/2$.
The first one has a pretzel-like shape and the second looks more like a fish.
The pretzels are usually flat and much more abundant than the fish, which are vertically extended.

In row (f) we plot an $\omega_y/\omega_x\approx 5/3$ orbit.
Commonly, these orbits are flat and have a clear boxy outline. The ratio of their sizes in the $x$ and $y$ directions
corresponds to the axis ratio of the bar itself. In row (g) we show a non-periodic box orbit, more elongated than
the previous one.

Here we have shown mostly (almost) periodic orbits, however, one should remember these are exceptions rather than the
typical cases. Most of the orbits do not have frequency ratios that can be approximated as ratios of small integers
and are not periodic. However, their outlines are related to the closest periodic orbits.

\begin{figure}
\centering
\includegraphics[width=\columnwidth]{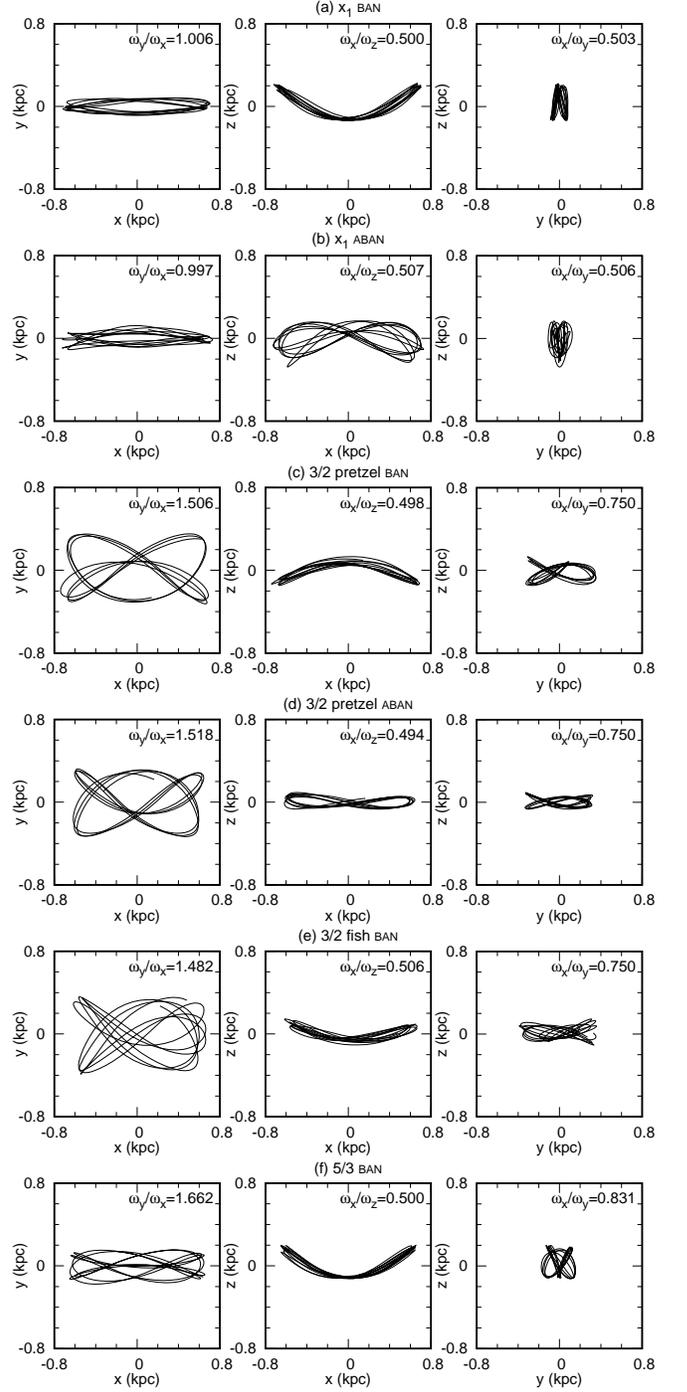}
\caption{Face-on, edge-on and end-on views of orbits which are in the $\omega_x/\omega_z\approx 0.5$ vertical
resonance and have {\sc (a)ban} shapes in the edge-on view. The corresponding frequency ratios are indicated in the
panels. All these orbits have the average size of $\bar{r}=0.5$ kpc.}
\label{fig_orbits_ban}
\end{figure}

In Figure \ref{fig_orbits_ban} we plot orbits which are in the $\omega_x/\omega_z\approx 1/2$ vertical resonance.
In row (a) we show a vertical bifurcation of the x$_1$ orbit, called x$_1$v$_1$ in the notation of \citet{skokos02}.
It was noticed for the first time by \citet{pfenniger_friedli91}, who called it {\sc ban}.
In row (b) we give another example of vertically extended counterpart of x$_1$ orbit, called x$_1$v$_2$ or {\sc aban}. \citet{pfenniger_friedli91} found
in their simulation
a whole sequence, from banana-shaped {\sc ban}s to infinity symbol-shaped {\sc aban}s. In our simulation
we found all of the vertical resonances of x$_1$ orbit of this kind. However, we found significantly more {\sc ban}s than {\sc aban}s.

In  rows (c) and (d) we present two 3/2 pretzel orbits in the $1/2$ vertical resonance. The first one is a {\sc ban}
and the second one is an {\sc aban}. A substantial fraction of 3/2 orbits is found in this resonance. Additionally, in
row (e) we show a {\sc ban} 3/2 fish orbit.
We complete Figure \ref{fig_orbits_ban} with row (f), depicting a 5/3 orbit, which is also banana-shaped in the $xz$
view. However, this type of orbit rarely exhibits this resonance.

\begin{figure}
\centering
\includegraphics[width=\columnwidth]{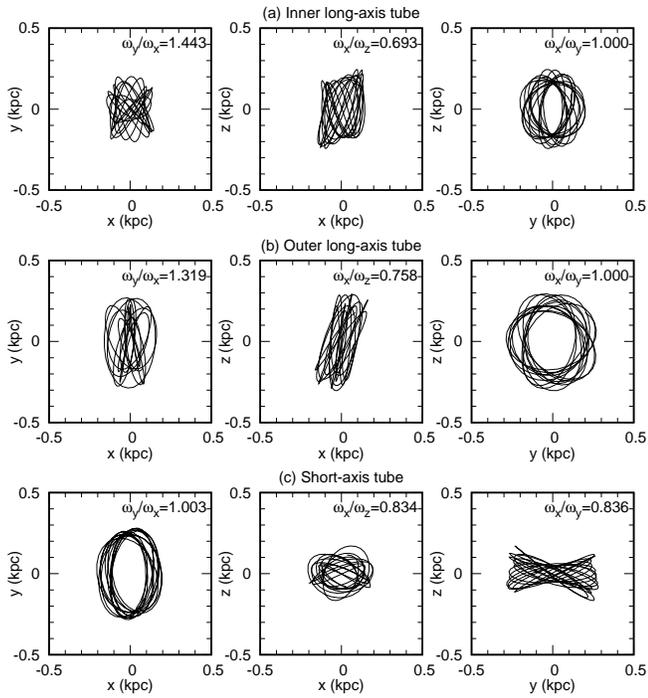}
\caption{Face-on, edge-on and end-on views of orbits known to exist in a triaxial potential. The corresponding
frequency ratios are indicated in the panels. All these orbits have the average size of $\bar{r}=0.2$ kpc.
}
\label{fig_orbits_ell}
\end{figure}

In Figure \ref{fig_orbits_ell} we display orbits present in the very central part of the bar.
They have the same shapes as the orbits in triaxial potentials, described by \citet{dezeeuw85}.
Rows (a) and (b) of the Figure show long-axis tube orbits, i.e. revolving around the longest axis of the bar.
The one in row (a) is the inner long-axis tube, distinguished by its extension along the $x$ axis.
The outer long-axis tube in row (b) is thinner in the $x$ direction.
It is worth noticing that they are not symmetric in the $xz$ view due to the Coriolis force.
Such a behavior was also noticed in the case of orbits in a rotating triaxial potential \citep{heisler82}.
In row (c) we present a short-axis tube which revolves around the shortest axis of the bar.
We note that all orbits of this kind are retrograde, thus in the $xy$ plane they rotate in the opposite direction
to the bar itself. In rotating triaxial potentials the retrograde short-axis tubes also replace the prograde ones
\citep{deibel11}. Orbits of shapes similar to the ones in the triaxial potential are present only in the very center of
the bar, presumably because the potential there resembles the one of the rotating ellipsoid.

\subsection{Distributions of frequency ratios}

We divided the particles according to their mean distance from the center of the dwarf, $\bar{r}$. Centers of the bins
span the range from $\bar{r}=0.2$ kpc to $\bar{r}=0.8$ kpc and have equal widths of $0.1$ kpc. The position of the bin
with the smallest $\bar{r}$ was set by the resolution of the simulation, whereas the largest $\bar{r}$ was chosen
so as to ensure a reliable frequency identification of close peaks in the $y$ spectrum, as we already discussed. In each bin we measured the frequencies for $5 \times 10^4$
particles.

We note that the mean distance of a given particle from the bar center may be significantly
smaller than its maximum extent, especially along the bar major axis. Particles in the
$\bar{r}=0.2$ kpc bin may well reach $x=0.3$ kpc. The orbits with $\bar{r}=0.5$ kpc extend to about
$0.7$ kpc in the $x$ direction, as can be confirmed by referring to Figure \ref{fig_orbits_flat}.
Moreover, the orbits with $\bar{r}=0.8$ reach approximately $x=1.1$ kpc. We estimated the bar
length to be approximately $1.8$ kpc, hence in our analysis we fully cover the inner 1/2 of the bar
and have strong indications for up to 2/3.

Similar fraction of the bar content was covered in terms of particle numbers. There are about $0.18\times 10^5$
particles with $\bar{r}<0.15$ kpc, i.e. closer to the center than the innermost bin. In the region we analyzed, i.e.
$0.15< \bar{r} < 0.85$ kpc, there are $5.22\times 10^5$ particles. In the range $0.85<\bar{r}<1.8$ kpc, namely in the
outer part of the bar, not studied in this work, there are $2.10\times 10^5$ particles. Hence, we studied the
particles comprising about $70\%$ of the stellar mass contained in the bar.

\begin{figure}
\centering
\includegraphics[width=\columnwidth]{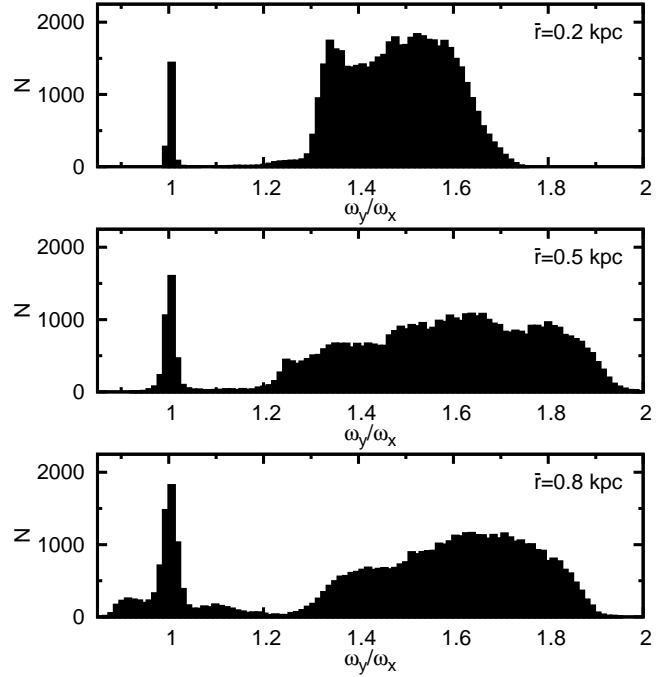}
\caption{Histograms of the orbital frequency ratio $\omega_y/\omega_x$. The three panels, from top to bottom, show
results for bins of $\bar{r}=0.2$, $0.5$, $0.8$ kpc, respectively.}
\label{fig_histogram_yx}
\end{figure}

In Figure \ref{fig_histogram_yx} we plot histograms of the ratio $\omega_y/\omega_x$ for three bins of $\bar{r}$.
The peak around $\omega_y/\omega_x\approx 1$ corresponds mostly to x$_1$-type orbits, but includes also short-axis
tubes in the innermost bin. The height of the peak grows with the mean radius, which means that the fraction of the
x$_1$ orbits grows as well. The increase of the width of the peak is related to our method of retrieving frequencies,
which is more accurate in the center of the dwarf.

The wide distribution of frequencies in the range $1.2$-$1.9$ changes significantly with radius. As can be judged
from Figure \ref{fig_orbits_flat}, the orbits with larger $\omega_y/\omega_x$ are horizontally thinner in the face-one view.
Thus, the orbits further from the center of the bar are on average more elongated than the ones in
the center.

In the innermost bin, besides the x$_1$ peak at $\omega_y/\omega_x=1$, there are two noticeable components: one
around $\omega_y/\omega_x\approx 1.33$ and the other, very broad, at $\omega_y/\omega_x\approx 1.5$. Thus, the
particles in the center are not on very elongated orbits, but their orbits are more square-like. The distribution at
$\bar{r}=0.5$ kpc resembles most the overall distribution of the particle orbits discussed in \citet{gajda_proc15}. It
is worth noting that the most represented orbits are the 5/3 ones. The orbits with $\omega_y/\omega_x\approx 1.9$ are
almost as elongated as the x$_1$ family.  In the outermost bin there are two additional small peaks in the
vicinity of $\omega_y/\omega_x=1$. We investigated the shapes and subdominant frequencies of the orbits belonging to
them. The peak around $\omega_y/\omega_x\approx 0.9$ consists of orbits similar to the one in Figure
\ref{fig_orbits_flat} (c), which has two comparable peaks at $\approx \omega_x$ and $\approx 1.3\omega_x$ in the $y$
spectrum. The other group of particles, at $\omega_y/\omega_x\approx 1.1$, is mostly made of orbits perpendicular the
bar (akin to Figure \ref{fig_orbits_flat} (b)), but we cannot exclude the possibility that some of them actually belong
to the disk population.

\begin{figure}
\centering
\includegraphics[width=\columnwidth]{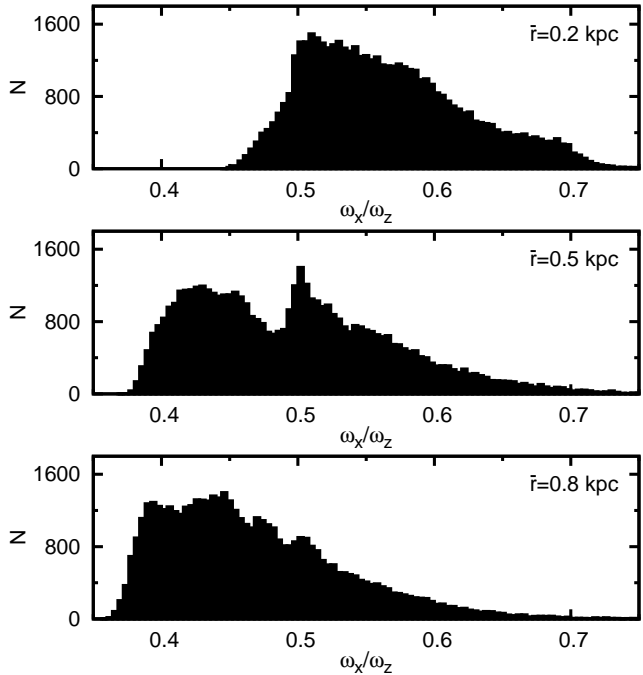}
\caption{Histograms of orbital frequency ratio $\omega_x/\omega_z$. The three panels, from top to bottom, show
results for bins of $\bar{r}=0.2$, $0.5$, $0.8$ kpc, respectively.}
\label{fig_histogram_xz}
\end{figure}

In Figure \ref{fig_histogram_xz} we show histograms of the $\omega_x/\omega_z$ frequency ratios. Let us note a trend that the orbits of the particles with $\omega_x/\omega_z>0.5$ are on average vertically thicker than the orbits having $\omega_x/\omega_z<0.5$.
Obviously, those which have $\omega_x/\omega_z=0.5$ are in the {\sc (a)ban} resonance.

In the innermost bin most of the orbits are vertically extended. In the intermediate bin more than half of the
particles are on rather flat orbits, but there is also a large fraction of thick orbits. There is also a pronounced peak
at $\omega_x/\omega_z=0.5$, consisting of {\sc (a)ban} orbits. In the outermost bin there are fewer thick orbits and the
{\sc (a)ban} peak is barely visible. Most of the orbits have $\omega_x/\omega_z<0.5$, so they are flat.
In comparison to the intermediate bin, there are more particles with $\omega_x/\omega_z<0.4$.

\begin{figure}
\centering
\includegraphics[width=\columnwidth]{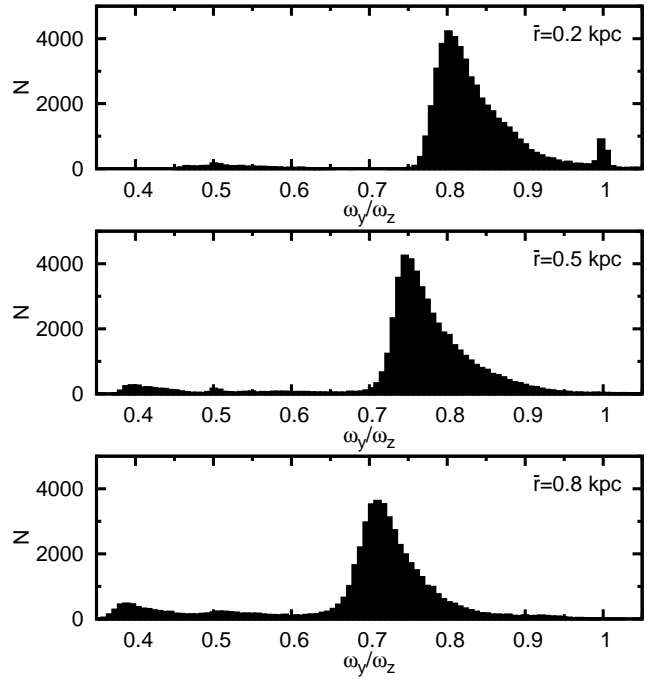}
\caption{Histograms of orbital frequency ratio $\omega_y/\omega_z$. The three panels, from top to bottom, show
results for bins of $\bar{r}=0.2$, $0.5$, $0.8$ kpc, respectively.}
\label{fig_histogram_yz}
\end{figure}

In Figure \ref{fig_histogram_yz} we show histograms of the $\omega_y/\omega_z$ frequency ratio. The overwhelming
majority of the particles is concentrated in one peak, which is asymmetric with respect to its maximum. However, the peak has different positions in different bins.
Its maximum is located at $\omega_y/\omega_z\approx 0.8$ in the innermost bin and at $\omega_y/\omega_z\approx 0.7$ in
the outermost bin. The small peak at $\omega_y/\omega_z\approx 1$ in the innermost bin consists of the long-axis tube
orbits (see Figure~\ref{fig_orbits_ell}). The orbits around $\omega_y/\omega_z\approx$ 0.4-0.5 are orbits with
$\omega_y/\omega_x \approx 1 $ (e.g. x$_1$) that are flat (i.e. $\omega_x/\omega_z < 0.5$). Both these conditions
result in $\omega_y/\omega_z < 0.5$.

\subsection{Orbit classification}

In this section we divide the particle orbits into families, differing in their shapes and contributions to the density
distribution. We identified five types of orbits: box, x$_1$, not supporting the bar, long-axis tubes and short-axis
tubes. Below we describe the criteria employed in this classification.

The largest fraction of \emph{x$_1$ orbits} are simple loop orbits (Figure~\ref{fig_orbits_flat}a), with
$\omega_y/\omega_x=1\pm0.035$. However, as predicted by theoretical studies
\citep[e.g.][]{contopoulos_papayannopoulos80}, some x$_1$ orbits have small loops at their ends. If the loops are large
enough, a different frequency dominates the $y$ spectrum, therefore we also add to this category particles with
$\omega_y/\omega_x=3\pm0.035$. Unfortunately, in many cases only one loop is well developed, so in order to classify
these orbits we need to add more elaborate criteria: $\omega_y/\omega_x=2\pm0.1$ and the second and the third frequency
in the $y$ spectrum being equal to either $\omega_x$ or $3\omega_x$. The x$_1$ orbits in our simulation are very elongated,
so we also put a constraint on the amplitude ratio: $A_y/A_x<0.15$.

The family of \emph{orbits not supporting the bar} contains particles admitting $\kappa/(\omega_x+\omega_y)=1\pm0.05$,
which usually have shapes similar to the ones in Figure~\ref{fig_orbits_flat} (b) and (c). We additionally include here
orbits that are elongated in the direction perpendicular to the bar (i.e. $A_y>A_x$) and not classified as tube orbits.
Moreover, orbits having $\omega_y/\omega_x<1.2$ and not assigned elsewhere were also assigned to this group.

The family of \emph{long-axis tube orbits} is populated with particles revolving around the $x$-axis, i.e. the long
axis of the bar, which are depicted in Figure~\ref{fig_orbits_ell} (a) and (b). This condition is realized by setting
$\omega_y/\omega_z=1\pm0.025$. We are not able to differentiate between the inner and outer subtypes, partially due to
the asymmetry in the edge-on view.

Particles on \emph{short-axis tube orbits} revolve around $z$-axis, i.e. the shortest axis of the bar. Thus, they are
characterized by $\omega_x/\omega_y=1\pm 0.05$, a condition almost the same as for x$_1$ orbits. However, they are
perpendicular to the bar, so we impose $A_y>A_x$. Additionally, after inspecting their frequency ratios in the plane
of the disk, we added another constraint $0.75 < \omega_x/\omega_z<0.95$.

All of the remaining orbits were classified as \emph{box orbits}. Due to the exhaustive constraints adopted for previous
families, only orbits depicted in Figure~\ref{fig_orbits_flat} (d)-(g), belong to this family.

\begin{figure}
\centering
\includegraphics[width=0.85\columnwidth]{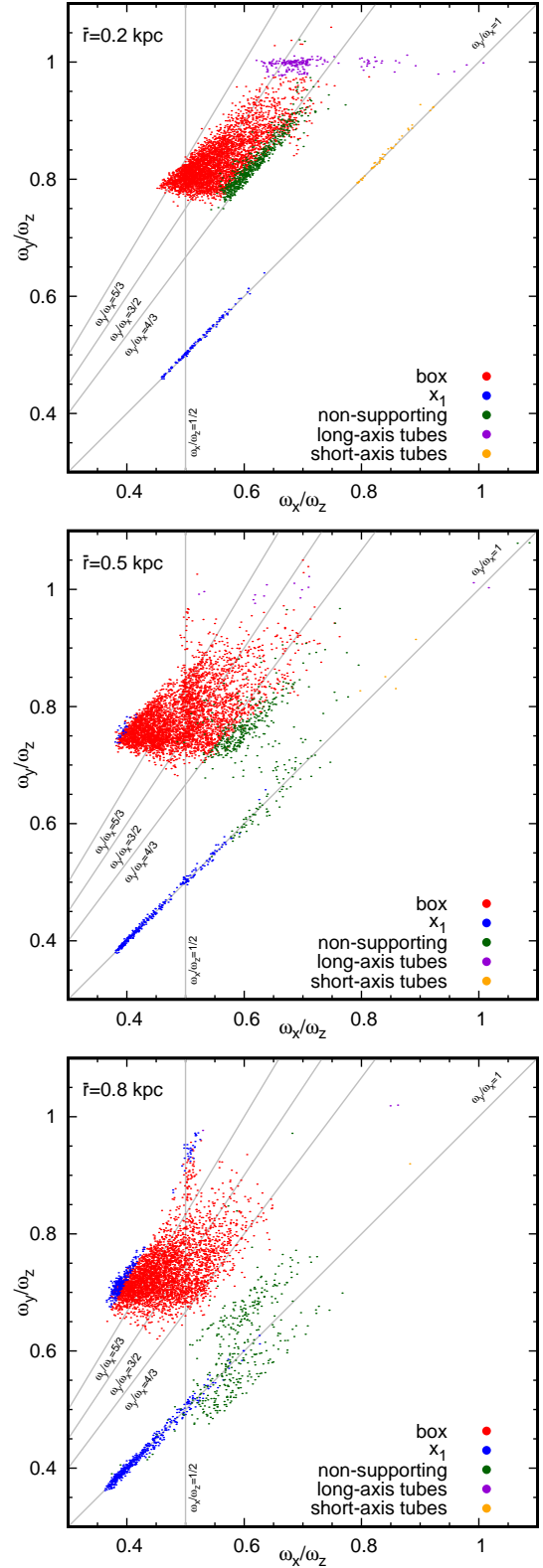}
\caption{Maps of the particle frequency ratios. Different colors indicate different orbit families: box (red), x$_1$
(blue), not supporting the bar (green), long-axis tubes (purple) and short-axis tubes (orange). The grey lines were
added to guide the eye: $\omega_y/\omega_x=1$ and $\omega_x/\omega_z=1/2$, as well as
$\omega_y/\omega_x=5/3$, $3/2$ and $4/3$. The three panels, from top to bottom,
show results for bins of $\bar{r}=0.2$, $0.5$, $0.8$ kpc, respectively.}
\label{fig_frequency_maps}
\end{figure}

In Figure~\ref{fig_frequency_maps} we present frequency ratios in the $(\omega_x/\omega_z ,\ \omega_y/\omega_z)$ plane.
Particles belonging to each orbit family are marked with different colors. In order to avoid overcrowding, we plotted
only $5\times 10^3$ particles in each panel. To guide the eye, we added grey lines corresponding to selected important
frequency ratios: $\omega_y/\omega_x=1$ and $\omega_z/\omega_x=1/2$, as well three others (from
top to bottom): $\omega_y/\omega_x=5/3$, $3/2$ and $4/3$.

First of all, it can be immediately noticed that the tube orbits (purple and orange) are present in large numbers only
in the innermost bin and there are only very few of them in the outer bins.
The x$_1$ orbits (blue) are well represented in all bins, but they become more frequent as the radius grows.
Their vertical thickness is also correlated with the distance from the center.
In the outer parts they are flatter than in the inner parts, as indicated by their $\omega_x/\omega_z$.
A part of x$_1$ orbits distribution overlaps with the vertical $\omega_x/\omega_z=1/2$ line at all radii, indicating
that x$_1$v$_1$ ({\sc ban}) and x$_1$v$_2$ ({\sc aban}) orbits exist throughout the whole bar. Even some clustering of
the dots at this resonance can be seen in the intermediate bin. In the innermost bin, the x$_1$ orbits are clearly
separated from the short-axis tube orbits. The width of the population with respect to the $\omega_y/\omega_x=1$ line
is related to the uncertainty of the frequency determination. As already discussed, our method is more accurate in the
inner part than in the outer part of the bar.

The orbits which do not support the bar (green) are located at $\omega_y/\omega_x \lesssim 1.33 $. Such a frequency
ratio is not surprising, as those orbits are not as elongated as the bar. Moreover, they are vertically thick, as
indicated by $\omega_x/\omega_z \gtrsim 0.6$. In the outermost bin, the huge spread of these orbits is probably
due to the misidentification of the frequency with the largest amplitude in the $y$ spectrum (the time period
considered was too short in comparison with the orbital period). On the other hand, those might also be
particles belonging to the disk population.

The box orbits (red) constitute the majority of the stellar particle orbits in the bar. The area they occupy has very
well defined boundaries, especially in the innermost bin. In the outer ones, the boundaries are blurred due to the
reduced accuracy. The cloud of points representing the box orbits changes its position with radius. In the outer
parts they have, on average, smaller $\omega_y/\omega_z$ and $\omega_x/\omega_z$ than in the inner parts. It
means that in the outer parts they are thinner and flatter.

In all bins, some of the box orbits are located on the $\omega_x/\omega_z=1/2$ line, so they exhibit {\sc
(a)ban} shapes. Depending on the distance from the center, orbits with all possible $\omega_y/\omega_x$ ratios may be
in this vertical resonance, especially in the range $\omega_y/\omega_x \approx 1.50-1.67$. Examples of these orbits were
presented in Figure~\ref{fig_orbits_ban} (c)-(f). In the intermediate bin, even some clustering of points close to the
$\omega_x/\omega_z=1/2$ line is visible.

\begin{figure}
\centering
\includegraphics[width=\columnwidth]{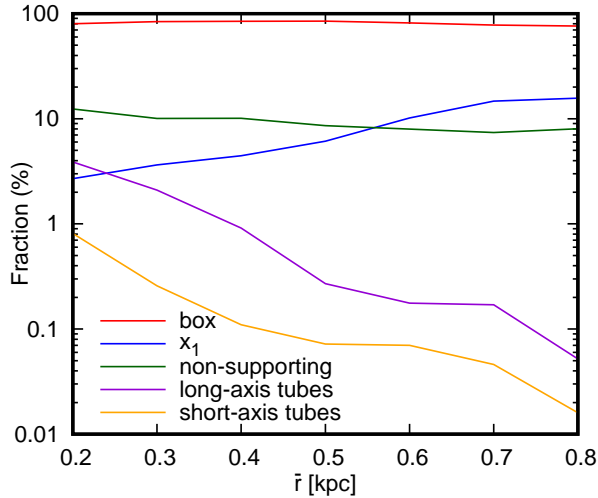}
\caption{Fractions of orbit families as a function of the distance from the dwarf center. The colors are the same as in
Figure~\ref{fig_frequency_maps}: box orbits in red, x$_1$ in blue, not supporting the bar in green, long-axis tubes in
purple and short-axis tubes in orange.}
\label{fig_classification_radial}
\end{figure}

In Figure \ref{fig_classification_radial} we show how the contribution of different orbit families varies with the
distance from the bar center. As already pointed out, the box orbits are the most prevalent.
Their contribution varies between $76\%$ and $85\%$ of all the orbits and shows only a small decrease with the distance from the center. The x$_1$ orbits
are quite uncommon at the center (less than $3\%$), but their fraction grows with radius, up to $16\%$ at
$\bar{r}=0.8$ kpc. The contribution of orbits not supporting the bar is about $10\%$ and slowly decreases with
radius. The orbits known to exist in triaxial potentials (short- and long-axis tubes) make up a few percent of the
distribution and practically disappear beyond $\bar{r}=0.5$ kpc.

In order to obtain the total contributions of the different families, we need to weigh the results from
Figure~\ref{fig_classification_radial} by the total number of particles present in a given bin.
As a result, we obtain the following percentages: box orbits: $81.7\%$, x$_1$ orbits: $8.0\%$, orbits not supporting the
bar: $9.2\%$, long-axis tubes: $1.0\%$ and finally short-axis tubes: $0.2\%$  (percentages do not add up to $100\%$ because of rounding). It is difficult to give an estimate of the error bars for these fractions, particularly because the criteria we applied (e.g. for the frequency or amplitude ratios) can only be given approximately, and because the box orbits category is composed of all orbits that do not belong to any of the other categories we have defined. Moreover, all the chaotic orbits are also included in the box category. Nevertheless, we believe that the general picture obtained from the numbers we give is correct.

\subsection{Buckling orbits}

\begin{figure}
\centering
\includegraphics[width=\columnwidth]{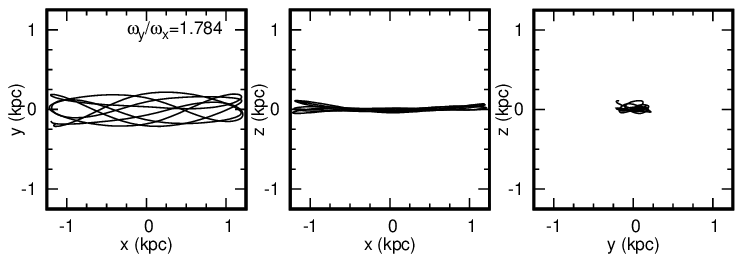} \\
\includegraphics[width=\columnwidth]{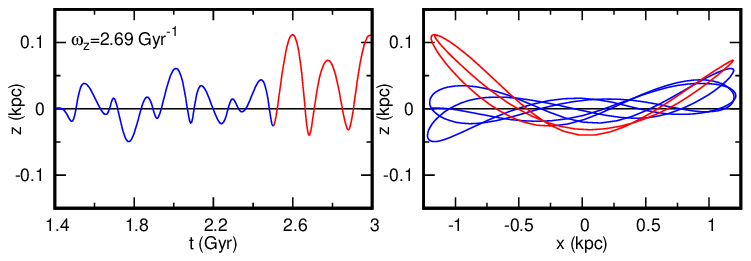}
\caption{Example of an orbit with a very small value of $\omega_z$. In the top row we show three projections of the
orbit: face-on, side-on and end-on (from the left to the right). The left panel of the bottom row plots the function
$z(t)$ for the orbit. At $2.5$ Gyr its character changes, which was emphasized by switching from the blue to the red
line. The right panel of the bottom row shows the projection of the orbit onto the $xz$ plane, but with the $z$
dimension enlarged. The color coding corresponds to the phases depicted in the $z(t)$ plot (blue for $t<2.5$ Gyr, red
for $t>2.5$ Gyr).}
\label{fig_buckling_orbit}
\end{figure}

For some of our particles we obtained exceptionally small values of $\omega_z$, of about $2.5$ Gyr$^{-1}$,
corresponding to periods of $2.5$ Gyr. We investigated why they seemed to have much longer vertical periods than
the periods along the major axis of the bar.
Figure~\ref{fig_buckling_orbit} shows an example of such an orbit. The panel plotting the $z(t)$ function explains
why we found such a low value of $\omega_z$. Before $t=2.5$ Gyr the particle was oscillating around $z=0$
with a small amplitude. After $2.5$ Gyr the amplitude grew significantly and in addition the oscillation became
asymmetric with respect to the $z=0$ plane. Simply, the change of the amplitude dominates the Fourier spectrum of
$z(t)$. In the close-up of the orbit projection onto the $xz$ plane, the change in the particle trajectory is also
well visible. Before $2.5$ Gyr (blue) the particle trajectory is simply flat but afterwards (red) it becomes a {\sc
ban} orbit.

The distribution of the particles with low $\omega_z$ is sharply peaked around $\omega_z=2.5$ Gyr$^{-1}$, and ends
by $\omega_z=4$ Gyr$^{-1}$, thus we use this value as a discriminant for this type of orbits. In the outermost radial
bin we have found that about $1.2\%$ of the particles exhibit this feature. Towards the center the fraction drops
quickly and they disappear completely from $\bar{r}=0.6$ kpc inward.

\begin{figure}
\centering
\includegraphics[width=\columnwidth]{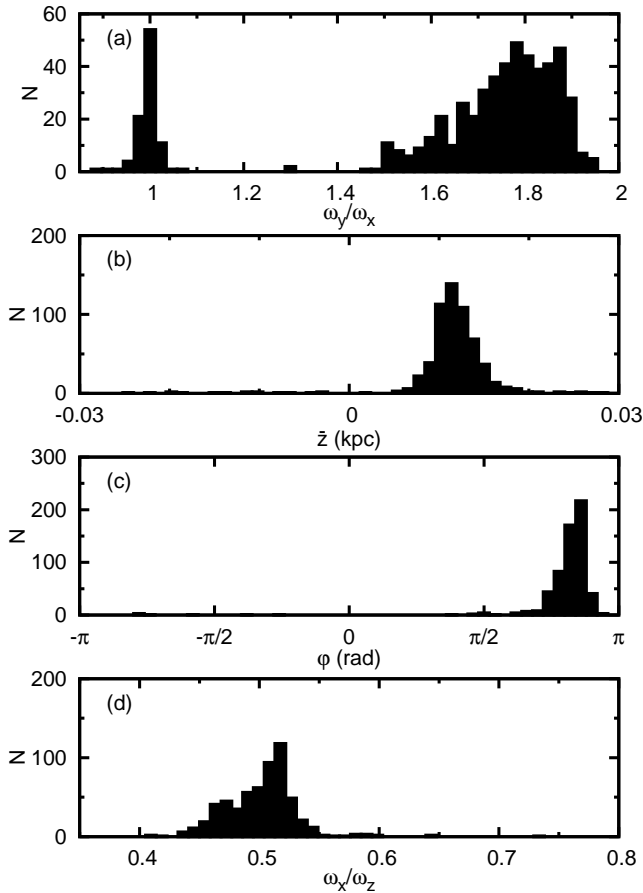}
\caption{Histograms describing orbits with $\omega_z<4$ Gyr$^{-1}$ and $\bar{r}=0.8$ kpc. Panel (a):
$\omega_y/\omega_x$ frequency ratio. Panel (b): the average of $z(t)$. Panel (c): phase of cosine wave $\cos(\omega_z t
+ \varphi)$ associated with the small $\omega_z$. Panel (d): $\omega_x/\omega_z$ frequency ratio, but in this panel $\omega_z$
is calculated from the $z(t)$ spectrum for the second half of the period of interest, $t \in (2.2, 3)$ Gyr.}
\label{fig_buckling_histograms}
\end{figure}

In Figure \ref{fig_buckling_histograms} we present distributions of various properties of the buckling orbits (i.e.
$\omega_z < 4$ Gyr$^{-1}$) from the outermost bin ($\bar{r}=0.8$ kpc). In the first panel we show the histogram of
$\omega_y/\omega_x$ frequency ratios. It turns out that predominantly the most elongated orbits buckle, either x$_1$ or
elongated boxes ($\omega_y/\omega_x\approx 1.8$). However, other types of box orbits, like 5/3 and 3/2 are also
represented.

In the second panel, we show the distribution of the mean of $z(t)$. Almost all the orbits have $\bar{z}>0$,
which means that they in fact do not stay in the $xy$ plane but above it. When the frequency is calculated, also
other parameters of the cosine wave $A\cos(\omega t+\varphi)$ are obtained, namely its amplitude $A$ and phase
$\varphi$. In the third panel we show the distribution of the phases of the small-frequency cosine wave. Almost all of
the particles have $\varphi \in (\pi /2, \pi)$, indicating that the outline of their oscillations grows with time.
Altogether, it means that the considered particles buckle upwards. As can be seen from the panels (b) and (c),
there are also a few particles whose orbits buckle downwards. To be sure what is the outcome of this process, we
calculated $\omega_z$, but only from the data for $t>2.2$ Gyr and showed the ratio $\omega_x/\omega_z$ in the last
panel. The ratios are grouped around $\omega_x/\omega_z\approx 1/2$, so the final products are of {\sc ban} type.

We note that here we study the bar before its main buckling event taking place around 3.5-3.8 Gyr \citep{lokas14}.
From the kinematics of the dwarf we infer that the behavior of these orbits might be regarded as a prelude to
the actual buckling, which is oriented downwards.

\section{Discussion}
\label{sec_discussion}
\subsection{Distribution of orbits}

It is essential to compare our results to the findings of other authors. Unfortunately, in
almost all other works the cylindrical frequencies were used to describe the orbits. Here we used
Cartesian frequencies, hence a direct comparison is not possible. From our limited experience with
the cylindrical coordinates we are only able to say that the $(\Omega-\Omega_p)/\kappa=1/2$ peak
seems to have the largest contribution to the distribution of orbits in the tidally induced bar,
similarly to what was found for the bars formed in isolation \citep{athanassoula02, athanassoula03,
martinez_valpuesta06}.

\citet{portail15} used Cartesian frequencies, but only to study the vertical structure. The only other work on $N$-body bars, where the full frequency analysis was performed in
Cartesian coordinates, is the one by \citet{valluri16}. However, \citet{valluri16} did
not study the real orbits of particles in their $N$-body simulation of an isolated disk galaxy, but
used the frozen potential method. They found that the bar is dominated by boxy orbits (almost
$70\%$). In addition, a significant amount (about $15\%$) of chaotic orbits was found to be
present. Added together, these values amount to approximately the same fraction of boxes as in our
simulation.

In their simulation \citet{valluri16} found only about $7 \%$ of x$_1$-shaped orbits. The fraction depends on the
radius and it is sharply peaked around half of the bar length. Here we found that the abundance of x$_1$ orbits grows
at least up to two-thirds of the bar length and presumably also further away. In the innermost part of the bar
\citet{valluri16} also found orbits known from triaxial potentials.

The frequency map of \citet{valluri16} also looks qualitatively similar to the ones shown in our
Figure~\ref{fig_frequency_maps} and the dependence of the position of the cloud containing the box orbits on the
distance from the center/Jacobi constant is also reproduced. The main difference between our results is the absence of
the diffuse cloud of particles around $\omega_y/\omega_x=1$. We believe that it might be the fault of our method which
is not able to properly assign $\omega_y$ frequencies to some orbits which are not elongated with the bar or may also
belong to the disk population or be chaotic (as those were not included in the frequency map in \citealt{valluri16}).

\citet{valluri16} treat the $\omega_y/\omega_x=3/2$ orbits, shown in our Figure~\ref{fig_orbits_flat} (d) and (e),
differently from the other boxes. However, in our simulation we did not find any significant clustering around this
particular resonance. Moreover, we have found more periodic box-like orbits: 4/3 shown in
Figure~\ref{fig_orbits_flat} (c) and 5/3 shown in Figure~\ref{fig_orbits_flat} (f).

Our study certainly has limitations, but it appears that the orbital structures of the tidally
induced bars and the bars formed in isolation are qualitatively similar to each other. We have found a few differences, however to understand and explain them, a deeper study would be needed.

\subsection{Periodic orbits and Surface of Section}

Another powerful tool, useful in studying the orbital structure, is the Surface of Section (SoS). These are
cuts through the phase space of solutions of a given Hamiltonian problem. If a particle is on a periodic orbit or is
trapped around one, its subsequent positions on SoS (called consequents) lie at the same point or form eventually a continuous curve. Moreover,
orbits trapped around different periodic orbits occupy a different part of the SoS. On the other hand, if the particle is on a chaotic orbit,
its consequents are scattered randomly. Of course, SoS does not tell the full story: the existence of a given orbit in
phase space does not mean that in reality or in a simulation this orbit would be populated by any objects. Here, we did not use SoS, however we would like to put our results in the context of previous studies.

\citet{athanassoula83} studied SoSs of two-dimensional analytic bars. They found that the size of the region related to
x$_1$ orbits (called family \emph{B} there) changes with the distance from the bar center. In fact, it grows with
the growing value of the Jacobi constant (which can be roughly translated to the distance from the center). This might
be related to the growth of the x$_1$ fraction with radius, which we find here. The x$_1$ island is surrounded by the sea of chaotic orbits.

\citet{voglis07} studied orbits in three-dimensional $N$-body simulations meant to represent barred disk galaxies. Among many orbits they found, the family S
resembles our orbit of Figure~\ref{fig_orbits_flat} (b) (albeit their example is symmetric with respect to the bar minor
axis). In the SoS it is placed near the region occupied by x$_2$ family of orbits that are perpendicular to the bar.
Their family R looks similar to our $\omega_y/\omega_x=4/3$ orbits shown in Figure~\ref{fig_orbits_flat} (c) and forms
a chain of stability islands around the x$_1$ region. The shape and position of their family P in the SoS seems to
indicate that it might be an intermediate family between the two mentioned before.

\citet{patsis_katsanikas14b} were concerned with the boxiness of isophotes of some galaxies and they searched for
orbits which may reinforce such shapes. They found some periodic orbits, which are also present in our simulation. Their
orbits rm21 and rm22 have the same shapes as our $\omega_y/\omega_x=3/2$ pretzel orbits shown in
Figure~\ref{fig_orbits_flat} (d). It seems that the fish version of the 3/2 orbit would occupy the same spots in the
SoS, hence it is appropriate to treat those orbits as one type. The \citet{patsis_katsanikas14b} tr1 orbit corresponds to our
5/3 orbit from Figure~\ref{fig_orbits_flat} (f). Both these families occupy small stability islands in the SoS in
the vicinity of the x$_1$ stability curves. It corresponds to the fact that in our simulation the exact resonance did
not attract many particles. Both families are claimed to be bifurcations of the x$_1$ family.
A bar formed in an $N$-body simulation, analyzed by \citet{shlosman99}, also exhibits complex structure of phase space. In his case the four islands corresponding to 3/2 orbits are much larger and the region trapped around x$_1$ orbits is significantly smaller. In line with our results, the 3/2 orbits are present at lower values of Jacobi constant (i.e. closer to the center) than the three islands of 5/3 orbits.

\citet{patsis_katsanikas14b} studied the behavior of particles perturbed from the periodic 3/2 islands, which lie amidst
the chaotic sea. The orbits turned out to be chaotic, however they were sticky to the x$_1$ region, as their consequents
remain in the vicinity of x$_1$ region for a few hundred of the x$_1$ periods. \citet{valluri16} also found that the 3/2
orbits form a ring or chain of islands in the SoS. Hence, they may be considered as viable building blocks of the
density distribution.
Unfortunately, \citet{patsis_katsanikas14b} did not extensively study the 5/3 family, which seems to be more abundant in
our model \citep[see][]{gajda_proc15}. It would be interesting to check if their perturbed versions are also sticky in
phase space.

\subsection{Vertical structure}

Studies of the vertically extended orbits are important, as the boxy/peanut (b/p) bulges are results of the bar buckling
\citep[for a review, see][]{athanassoula16}. \citet{martinez_valpuesta06} analyzed periodic orbits during buckling and claimed that banana-shaped
x$_1$v$_1$ orbits are responsible for the asymmetric shape. \citet{patsis_katsanikas14b} supposed that 3/2 orbits, when
seen edge-on, may contribute to the b/p shape. It turned out that in the relevant SoS the 3/2 orbits are sticky to the
x$_1$v$_1$ stability islands. In our pre-buckling bar we found many 3/2 orbits with a {\sc ban} shape. However, the
orbits which we caught buckling, have either x$_1$ or elongated boxy shapes. Nonetheless, during the proper buckling
period also the 3/2 and other orbits may buckle. On the other hand, \citet{portail15} claim that \emph{brezel} orbits contribute the most to b/p bulges. In our notation they have $\omega_x/\omega_z=0.6$ and the particular example of this orbit shown in \citet{portail15} is a distorted version of $\omega_y/\omega_x=3/2$ pretzel orbit.

It may be considered surprising that we find in our simulations orbits trapped around the
x$_1$v$_2$ periodic orbit (Figure~\ref{fig_orbits_ban}b), as \citet{skokos02} found this family to
be unstable. Indeed, we found significantly fewer infinity-symbol shaped x$_1$v$_2$ orbits than
banana shaped x$_1$v$_1$ orbits, and this may be related to instability. Moreover, the particles we
found trapped on x$_1$v$_2$ orbits change their sense of circulation on the $\infty$ symbol. There
are thus two possible explanations of the existence of orbits trapped around the x$_1$v$_2$ family.
They could be non-periodic sticky orbits in the neighborhood of x$_1$v$_2$, as discussed by
\citet{patsis_katsanikas14a}, albeit in the different context. Alternatively, the x$_1$v$_2$ orbits
may be more stable in simulation potentials which are more realistic than the Ferrer's bar
potentials (for a discussion of the inadequacies of the latter, see \citealt{athanassoula15} and
\citealt{athanassoula16}). In general, when studying the vertical $1/2$ resonance one should take
into account the whole {\sc ban}-{\sc aban} sequence \citep{pfenniger_friedli91}.

\section{Summary and conclusions}
\label{sec_summary}

We have studied the orbital structure of a tidally induced bar formed in a dwarf galaxy orbiting
around a Milky Way-like host. We have analyzed the dominant frequencies during the period
between the first and the second pericenter passage, when the bar is strongest, rotates steadily
and has not yet buckled. Since we studied the real stellar orbits in the simulation and not in a
frozen potential, our analysis is limited to the inner two-thirds of the bar. Our results
should be relevant for at least some of dwarf galaxies in the Local Group, for which there are
indications of the presence of a bar \citep{coleman07, munoz10, lokas10, lokas12}.

For several reasons, described in detail in this paper, we decided not to use the cylindrical frequencies, but to use the Cartesian ones instead. An important one amongst these reasons is that calculating $\Omega$ properly is a complex task, which is beyond the scope of this paper and will be explored in another work.

We have found a variety of particle orbits, including obviously the x$_1$ orbits with $\omega_y/\omega_x=1$. However,
many of the stellar orbits turned out to have boxy shapes with various degree of elongation. In addition, we found
boxy periodic orbits with $\omega_y/\omega_x=4/3,\ 3/2$ and $5/3$. The boxy orbits can be either flat or
not, and they can be in the vertical $\omega_x/\omega_z=1/2$ resonance, exhibiting banana or infinity-symbol shapes in the edge-on
view. In the very center of the bar we encountered the orbits known from triaxial potentials: long-axis tubes (inner
and outer) and short-axis tubes (only retrograde).

The distribution of the box orbit shapes depends on the distance from the center of the bar. The box orbits in the
inner parts are on average less elongated than in the outer parts. Also the thickness of the orbits
changes with radius. In the center they are mostly vertically extended, but they are thinner further away. The {\sc
(a)ban} orbits are present at all radii, but their abundance varies. The differences in the orbital shapes are probably
related to the changes of the potential in the bar. Although the variation of the axis ratio with radius is small
(Figure~\ref{fig_shape_x}), it seems to have sufficient impact on the orbits. In the frequency maps the box orbits are
grouped in clouds, whose position changes with the distance from the center.

Using the frequency ratios and other criteria we have divided the orbits into five groups: boxes, classical x$_1$,
orbits not supporting the bar, long-axis tubes and short-axis tubes. The tubes are present only in the center, where the
potential is most similar to the rotating triaxial ellipsoid. The fraction of x$_1$ orbits grows from $2.5\%$ in
the center to $15\%$ in the outermost bin we considered here which is between half and 2/3 of the bar length.
However, in total we
found only $8\%$ of classical x$_1$ orbits. The most common ($80\%$) orbits in our tidally induced bar are boxes.
Hints concerning that can be already found in \citet{pfenniger_friedli91}. In their simulated bar $81 \%$ of
particles  were on orbits with an alternating direction of rotation. The full distribution of the $\omega_y/\omega_z$
ratio is peaked around the 5/3 periodic orbits, which have a similar shape in the face-on view to the bar $b/a$ axis
ratio.

We cannot compare our results directly to most of the previous ones, which used cylindrical
frequencies. However, our results are in broad agreement with the findings of \citet{valluri16},
obtained for a bar formed in an isolated Milky Way-like galaxy. Hence, it seems that despite
the different formation mechanism, the orbital structures are similar.

Even though we studied the bar before its main buckling event, we have found some orbits which buckle upward in the
outer part of the bar. It is worth noting that the main buckling event is most pronounced in the downward direction and
here we probably encountered the beginning of this process. The buckling orbits are either of the x$_1$ type or very
elongated boxes. Our method of searching for particles with exceptionally small values of $\omega_z$ seems to be a
promising tool for studying how orbits actually evolve during the buckling instability.

Boxy orbits seem to be an indispensable element of the density distribution in any barred galaxy. Therefore, more
work is needed regarding the parent, periodic orbits. When perturbed, they become chaotic but remain in the
vicinity of the stability islands, i.e. they are sticky. The distribution of frequency ratios of the box
orbits and their shapes may be influenced by the presence of various stability islands in phase space. Therefore,
comprehensive studies of phase space of not artificial, but realistic barred potentials would be very insightful.

\acknowledgments
\section*{Acknowledgments}
This work was partially supported by the Polish National Science Centre under grant 2013/10/A/ST9/00023.

\end{document}